\documentclass[a4page,onecolumn,11pt]{IEEEtran}
\usepackage{times}
\usepackage{cite}
\usepackage{color}
\usepackage{epsf}
\usepackage{epsfig}
\usepackage{graphicx}
\usepackage{graphics}
\usepackage{epstopdf}
\usepackage[small]{caption}
\usepackage{amsmath}
\usepackage{amssymb}
\usepackage{amsthm}
\usepackage{amsxtra}
\usepackage[ruled]{algorithm2e}
\usepackage{algorithmic}
\usepackage{enumerate}
\usepackage{multirow}
\usepackage{enumerate}
\usepackage{amssymb}
\usepackage{lipsum}
\usepackage{setspace}
\usepackage{multirow}
\usepackage{subcaption}
\usepackage{url}
\usepackage{lineno}
\usepackage{rotating} % adde by Wayes Tushar
\usepackage{graphicx} % added by Wayes Tushar
\newcommand{\Rmnum}[1]{\expandafter\@slowromancap\romannumeral #1@}

\makeatother
\begin{document}
%\setpagewiselinenumbers
%\modulolinenumbers[1]
%\linenumbers
%
\title{IoT for Green Building Management}
\author{Wayes~Tushar,~\IEEEmembership{Senior Member,~IEEE,}~Nipun Wijerathne,~Wen-Tai Li,~\IEEEmembership{Member,~IEEE,}~Chau~Yuen,~\IEEEmembership{Senior Member,~IEEE,}~H. Vincent Poor,~\IEEEmembership{Fellow,~IEEE},~Tapan Kumar Saha,~\IEEEmembership{Senior Member,~IEEE,} and~Kristin L. Wood
\thanks{W. Tushar and T. K. Saha are with the School of Information Technology and Electrical Engineering of the University of Queensland, QLD, Australia. (Email: wayes.tushar.t@ieee.org; saha@itee.uq.edu.au).}
\thanks{N. Wijerathne is with the Cloud Solutions International (Pvt) Ltd, Sri Lanka. (Email: hnipun@gmail.com).}
\thanks{W.-T. Li, C. Yuen, and K. L. Wood are with the Singapore University of Technology and Design (SUTD), 8 Somapah Road, Singapore 487372. (Email:\{wentai\_li, yuenchau, kristinwood\}@sutd.edu.sg).}
\thanks{H. V. Poor is with the Department of Electrical Engineering of Princeton University, Princeton, NJ 08544, USA. (Email: poor@princeton.edu).}
}\IEEEoverridecommandlockouts
\maketitle\doublespace
%\thispagestyle{fancy}
% ================================
\begin{abstract}
Buildings consume $60\%$ of global electricity.  However,  current building management systems (BMSs)  are highly expensive and difficult to justify for small to medium-sized buildings. As such, the Internet of Things (IoT), which can monitor and collect a large amount of data on different contexts of a building and feed the data to the processor of the BMS, provides a new opportunity to integrate intelligence into the BMS to monitor and manage the energy consumption of the building in a cost-effective manner.  Although an extensive literature is available on IoT based BMS and applications of signal processing techniques for some aspects of building energy management separately, detailed study on their integration to address the overall BMS  is quite limited.  As such, the proposed paper will address this gap by providing an overview of an IoT based BMS leveraging signal processing and machine learning techniques. It is demonstrated how to extract high-level building occupancy information through simple and low-cost IoT sensors and studied the impact of human activities on energy usage of a building, which can be exploited to design energy conservation measures to reduce the building's energy consumption.
\end{abstract}
% =================================
 \setcounter{page}{1}
 % =======================================================
\section{Introduction}\label{sec:Introduction}
% ========================================================
Buildings are one of the major electricity consumers being responsible for $60\%$ of total global electricity consumption. In the United States, for example, $70\%$ of annual electricity use is due to buildings~\cite{Tushar_TCSS_2018}. Such intense electricity usage by buildings is also true for many other countries, while detailed statistics may not fully available due to the lack of information or measurement. Therefore, there has been a significant push towards studying and developing means to effectively manage electricity in buildings through efficient building management systems (BMSs).

Current BMS solutions are, however, highly expensive and have been proven to be difficult to justify for use in small and medium-size buildings. Additionally, due to a recent push for reducing electricity consumption and increasing operational efficiency, building managers need to deal with dynamic and diverse requirements of buildings including anomaly detection, predictive maintenance, occupancy tracking, and electricity usage optimization with renewable integration. For example, in the US the heating, ventilation, and air conditioning (HVAC) airflow is regulated by OSHA (Occupational Safety and Health Administration) and is tied to maximum occupancy. Consequently, if no sensors are present, a room must be ventilated during normal working hours according to the maximum number of people who could be in the room (maximum seating capacity), thereby wasting a lot of energy. Indeed, sensing capabilities can lead to better situational awareness and more efficient and more dynamic and adaptive management of electricity and energy storage devices through incorporating intelligence into the BMS. As such, the Internet of Things (IoT) has emerged as a promising solution to make this integration a reality.
\begin{figure}[t]
\centering
\includegraphics[width=0.8\textwidth]{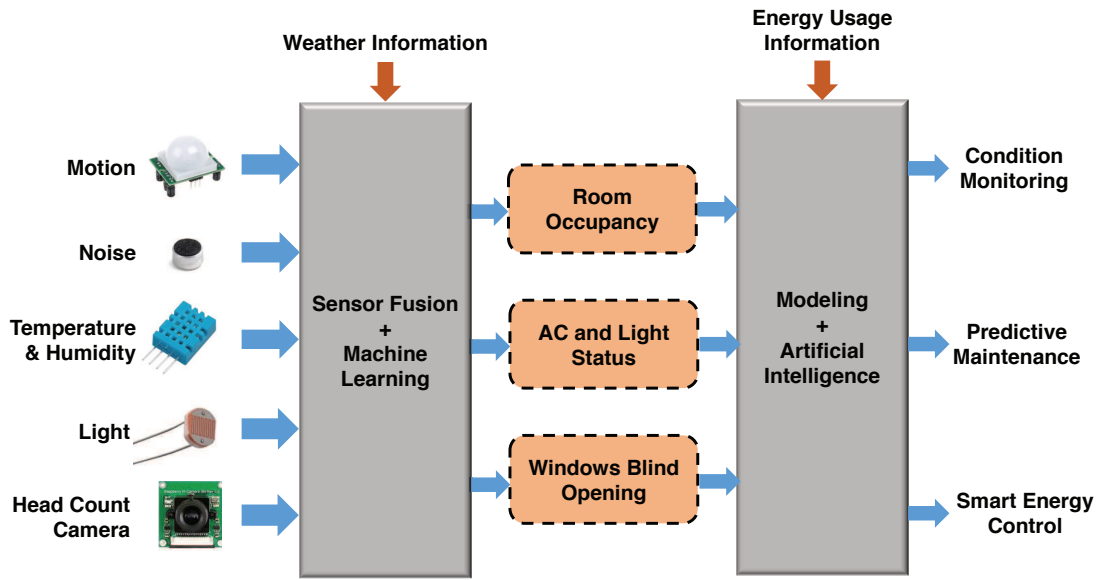}
\caption{Demonstration of the use of sensor fusion and machine learning for IoT based green building management.}
\label{fig:IoTBMS}
\end{figure}

Essentially, IoT is a platform that connects devices over the internet, lets them talk to each other and to humans, and by doing so enable the attainment of desirable context specific objectives such as energy savings (e.g., scheduling HVAC based on occupancy), condition monitoring (e.g., fault detection of HVAC), and predictive maintenance (e.g., the servicing of air filters in HVAC). As such, considering the fact that IoT is expected to change the future of smart BMS, this paper seeks to provide an overview of an IoT based BMS leveraging signal processing and machine learning techniques. We note that signal processing has been extensively used for wireless sensor networks for an assisted living and information filtering. Therefore, it could be very useful for extracting useful information about building health based on different sensors as depicted in Fig.~\ref{fig:IoTBMS}. From this perspective, we demonstrate an energy efficiency study that was conducted in a building testbed. The testbed is equipped with IoT devices and uses signal processing with machine learning to understand human activity and its impact on energy use of the building. To this end, the main contributions of this paper are in terms of
\begin{itemize}
\item Providing a literature review on the application of IoT in building management, and a discussion on the desired features of a smart BMS. 
\item Implementing machine learning techniques in low-level devices such as sensors and other IoT devices via transfer learning and semi-supervised learning techniques, to enable such IoT devices with low computation capability to perform machine learning algorithms locally via edge computation.
\item Illustrating how such techniques can benefit building management by providing useful information that can better characterize the energy efficiency.
\item Presenting some case studies from experiments in a real-world environment.
\end{itemize}

The rest of the paper is organized as follows. We provide an overview of the state-of-art of the application of IoT devices for BMS in Section~\ref{sec:LiteratureReview} followed by a description of trending BMS technologies in Section~\ref{sec:TrendingTechnologies}. A detailed discussion on low cost human activity detection with IoT devices in a selected space is provided in Section~\ref{sec:MultiSensorFusion} along with some experimental results. Finally, we provide some concluding remarks in Section~\ref{sec:Conclusion}.
% ============================================================
\section{State-of-the-Art}\label{sec:LiteratureReview}
As buildings undergo years of use, their thermal characteristics deteriorate, indoor spaces get rearranged, and usage patterns change. In time, their inner and outer microclimates adjust to the changes in surrounding buildings, overshadowing patterns, city climates and building retrofitting \cite{Manic_IEM_2016}. As a consequence, their performance frequently falls short of expectations. As such, IoT opens a new opportunity to integrate intelligence into the BMS in a cost-effective manner through seamless integration of various sensors, smart meters and actuators by which a BMS can monitor and identify different energy and environment-related parameters \cite{Pan_IoT_2015}, analyze the health of a building, determine energy and thermal requirements, and subsequently smartly determine the electricity usage behavior of different subsystems. Indeed, the performance of a BMS predominantly depends on the collection of large volumes of data from different subsystems of a building, which is then analyzed and processed by the BMS using various signal processing tools. Based on the application of IoT based signal processing techniques in managing various subsystems within a building, existing studies can be divided into five general categories: 1) lighting, 2) HVAC, 3) flexible load, 4) human detection, and 5) diagnostics and prognostics. These five categories are briefly described in the following paragraphs.

\subsection{Lighting}Lighting accounts for a major fraction of global electricity consumption. In office buildings, for example, the electricity consumed by lighting can be up to $40\%$ of the total electricity consumption~\cite{Pandharipande_ENB_2015}. From this perspective, a number of studies have been conducted to develop solutions that can help to reduce the electricity consumption by the lighting of buildings. For instance, with the purpose of achieving a desired illumination condition within a building with low electricity consumption, the authors in \cite{Pandharipande_ENB_2015} propose a luminaire-based periodic sensor processing algorithm in order to implement a smart lighting control system. In \cite{ZCheng_EB_2016}, the authors present a Q-learning based lighting control system that personalizes and employs users' perceptions of surroundings as the feedback signal for better management of lighting intensity. Further, \cite{Jain_BE_2018} and \cite{Guo_LRT_2010} present a review of the lighting control techniques by using signal processing based daylight prediction and occupancy detection methods respectively.
 
\subsection{HVAC}The HVAC system is another major consumer of electricity, which is responsible for $40\%$ of total electricity consumption of buildings in the United States~\cite{Tushar_TCSS_2018}. Consequently, developing means to reduce HVAC's electricity consumption has received considerable attention.  In \cite{Sun_TASE_2014}, the authors propose a Kalman filtering based gray box model to predict and determine statistical process control limits for fault detection of HVAC systems. Similar signal processing technique is also used in \cite{Ali_WPC_2013} for controlling the power consumption of buildings without compromising the comfort level of the occupants. An intelligent controller model is designed in \cite{Javed_TII_2017} by integrating IoT with cloud computing and Web services. Further, the authors develop wireless sensor nodes for monitoring the indoor environment and HVAC inlet air, and a wireless base station for controlling the actuators of HVAC. Lastly, a smart home energy management system using IoT and big data analytics is proposed in \cite{Ali_TCE_2017}, which predominantly focuses on the electricity consumption of the HVAC system. In particular, the proposed mechanism makes use of off-the-shelf business intelligence and big data analytics software packages to better manage energy consumption and to meet consumer demand. Other examples of such studies in the context of HVAC can be found in \cite{Aftab_EB_2017} and \cite{Mirakhorli_EB_2016}.

\subsection{Flexible loads}The third area of study focuses on monitoring and control of electricity consumption by other flexible loads within buildings. Example of such loads includes washing machines, dishwashers, ovens, electric vehicles, and energy storage systems. IoT devices can effectively monitor the operational status of these loads and exploit signal processing techniques to predict their usage patterns and effectively control their operation for better energy management (or demand response). For example, a learning-based signal processing tool for demand management is designed in \cite{Zhang_TSG_2016}. A deep learning based signal processing approach is implemented in \cite{Singh_TSG_2017} for non-intrusive monitoring of all loads in an entire building. In \cite{Kong_TPS_2017}, the authors design a long-short-term memory for load forecasting based on residential behavior learning using recurrent neural networks, and accurate indoor occupancy tracking within a building is implemented in \cite{Belmonte_Sensor_2017} using multi-sensor fusion. 

\subsection{Human detection}In the area of human detection for BMS, a number of machine learning techniques have been used for headcount and occupancy detection purposes. For example, parametric and non-parametric algorithms including background subtraction models and Gaussian processes \cite{LI_SPAC_2017} have been used with a camera for headcount. These algorithms are implemented using the OpenCV library. Further, for occupancy detection,  thermal imaging \cite{Tyndall_Sensor_2016}, pyroelectric infrared sensors \cite{Raykov_UbiCom_2016}, and RGB camera-based techniques have been used extensively.  An example of sound sensor-based applications for occupancy detection can also be found in  \cite{Salamon_ICASSP_2015} which uses high sampling rates for classification of activities. 

\subsection{Diagnostics and prognostics}The HVAC system is the most complex system and biggest energy consumer in buildings. Faulty equipment within HVAC leads to inefficient system operation. Hence, keeping HVAC systems is good operational condition is important. However, regular maintenance of HVAC system is time-consuming. Given this context, IoT devices allow us to develop advanced predictive maintenance, fault detection and diagnostics applications for the HVAC system. For instance, based on the data collected from IoT devices, a data-driven fault detection method is developed in \cite{Sun_TASE_2014}.

\begin{table}[t]
\centering
\caption{Summary of the surveyed literature on the application of IoT based signal processing for BMS.}
\includegraphics[width=\textwidth]{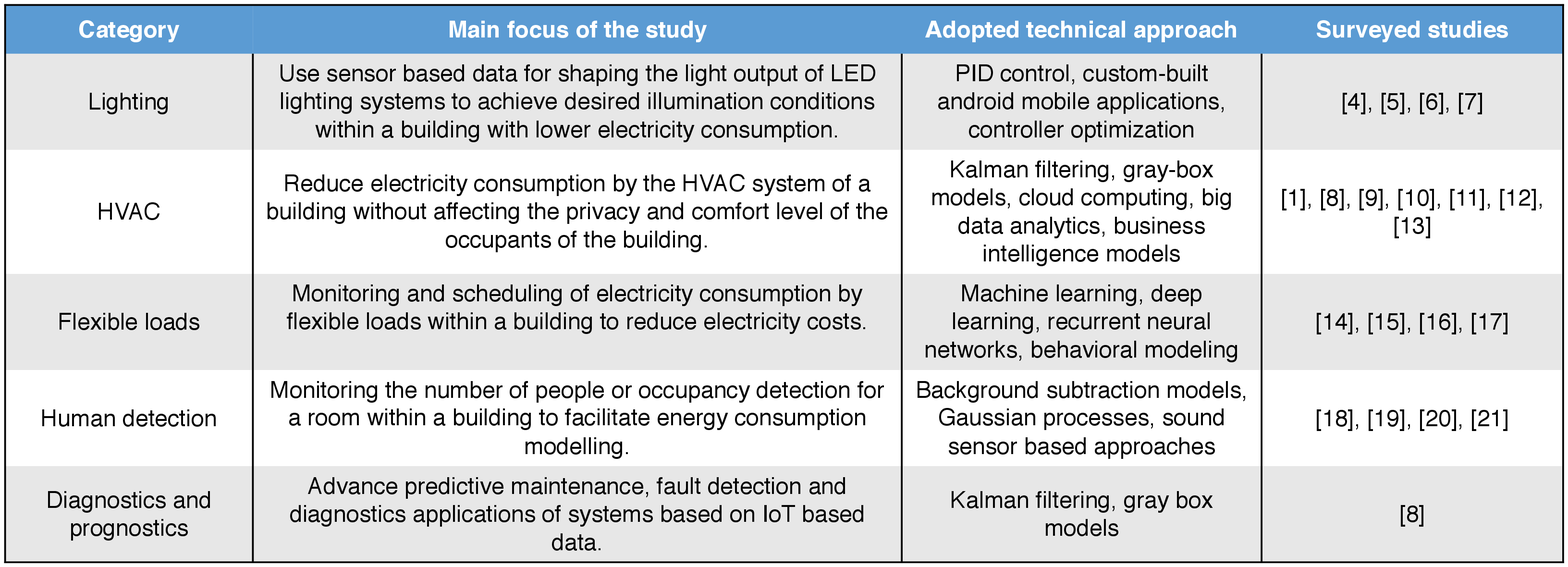}
\label{table:literature}
\end{table}

Now, based on the above discussion, which is also summarized in Table~\ref{table:literature}, clearly IoT has useful applications in designing smart and efficient buildings. Nevertheless, obtaining desirable performance of the BMS based on the applications of various signal processing techniques is highly dependent on the IoT devices that monitor and collect large amounts of data on respective contexts and then feed the data to the processor. Although studies are available on IoT based BMS and the applications of signal processing techniques for some aspects of building energy management separately (as noted above), studies on their integration to approach overall building energy operation in practical settings are limited. This paper addresses this gap by providing an overview of how IoT devices can be coupled with signal processing techniques to better understand a building's electricity usage performance. 

In the next section, first, we briefly discuss the trending technologies for BMS followed by a discussion on our experiments to be described in depth in Section~\ref{sec:MultiSensorFusion}.

% ============================================================
\section{Trending Technologies for BMS}\label{sec:TrendingTechnologies}
% ============================================================
This section provides an overview of the trending technologies that are being used for the effective design and development of data acquisition, management, and control platform of the BMS. Thus, these technologies prepare the BMS as a cloud-based ecosystem that 1) utilizes social interaction between the people within communities to establish sophisticated global behavior towards achieving different social, financial, and scientific goals; 2)  uses green energy and storage resources for environmental sustainability; and 3) overall affirms the establishment of a smart system with autonomous decision making capability.

\subsection{IoT for seamless integration and processing}
At present, it is considerably difficult and expensive for a building manager to be fully aware of the health of a building in real-time due to the limited sensing and control capabilities of current buildings. Hence, the design of an integrated data acquisition and control system based on an open architecture and cloud-enabled IoT can help to reduce the cost of having a BMS being set up. An integrated IoT system allows the building manager to monitor and sense different environmental parameters of the building (e.g., through motion and noise detectors, temperature and humidity sensors, and electricity and water flow meters), collect the relevant human activity information (occupancy, heat map), estimate the energy usage (e.g., by comparing the current information with previously collected historical data), which will be fed into a smart management system to manipulate actuators (e.g., switches, controllers and thermostats) to efficiently manage the building's environment according to expectations and designed rules.  

Such an IoT platform, which is open platform, can provide interface and connectivity with various subsystems of different vendors, e.g. sensing subsystem (people counting, temperature, humidity, light, noise, and motion), control subsystem (thermostats, switches, smart plugs, actuators), and metering subsystem (energy consumption, water flow). Currently, there are various off-the-shelf products and systems available in the market for IoT, e.g. there are trends of utilities worldwide towards the SEP (Smart Energy Profile) such as batteryless energy harvesting switch (Enocean), low-cost Wi-Fi controller (Particle), and Nest (by Google),  and many others. As the technology advances, it is expected to have more and more IoT devices to be available in the market. Hence there is great opportunity to tap the capability of these growing IoT devices. Nevertheless, for implementing an IoT system, one needs to address the challenges of scalability, flexible provisioning, interoperability and low latency~\cite{sanjana-IoT:2016}.
\subsection{Attribution of energy usage to human activities}
One of the key technologies that has received considerable attention for deploying as a part of BMS is the integration of human activity tracking within the control system so as to understand how electricity usage of a building is affected by its number of occupants and their various kind of activities. By tracking human activity, we refer to track human movements within a designated area, which can be done by using a smartphone scanning sensor. Essentially, a smartphone scanning sensor can scan the smart devices such as smartphones and mobile tablets in the vicinity, and at the same time, record the duration of stay and the MAC address of the smart device. This determines not only the heat map (i.e. human count) of the designated area but also manages to provide information on the duration of stay and movement path, and potentially the social relationship as well. Such tracking can provide building managers with information on efficient use of different areas in a building by the occupants and recommend if further modification (architectural or electrical system) of a space is necessary for becoming more energy efficient.

\subsection{Big data analysis for insightful analysis}
Big data management is essentially the core software layer that ultimately drives the BMS through big data analytics that includes prediction (e.g., predictive maintenance), model building, complexity mapping, and visualization. Big data management aids the dynamic management of the energy consumption via monitoring and analyzing the energy-related activities to minimize the unintended energy and water consumptions. It is essentially the process of examining the large data set obtained through the IoT. Through this process, the building manager is able to identify information on human activity, weather condition, the micro-climate within the building and related energy wastage. Indeed, to do so, algorithms need to be designed that can accurately extract the inter-correlation between the load consumption, human occupancy and movement activity, energy wastage, renewable energy generation, weather condition, and hence design effective model from real-time large-scale datasets in order to perform predictive maintenance.
\subsection{Renewable energy and storage for increasing flow of green energy}
To facilitate the flow and use of green energy, BMS also explore the provisioning of renewable energy resources in buildings through dynamic scheduling and control. In particular, first, the BMS collects data on weather conditions and subsequence renewable energy generation through the low-cost IoT system. Then, together with the information on human activity within the building, the BMS exploits its big data analytics to determine how to schedule the dispatch of the renewable energy from its distributed energy sources as well as the charging and discharging of respective storage devices. For example, solar thermal system integrated with hot water storage is becoming very popular in commercial buildings. Based on the above mentioned process, i.e., based on the available solar generation for heating water, demand of hot water based on human activity, the BMS system of a building can use its big data analytics to predict how much hot water will be needed for building. Thus, it can dynamically schedule heat pump to turn on to heat hot water based on the availability of renewable energy. 
% =======================================================
\section{IoT Based Human Activity Detection and Building Efficiency}\label{sec:MultiSensorFusion}
% =======================================================
As mentioned earlier, understanding human activity is of great importance to achieve energy efficiency. In this section, we provide a brief case study on how we can use low cost IoT devices together with signal processing and machine learning techniques to understand the number of people in a selected area of a building and provide the building management with key insights on how to effectively manage the electricity consumption of the building.

\subsection{Head counting with an overhead camera}To achieve a low-cost headcount camera, we use a camera with a fisheye lens that captures images at 30 frames per second. The camera is installed right on top of the entrance door of the selected room. The images captured by the camera is processed locally. Only the headcount number is uploaded to the cloud. For head counting, we explore a number of signal processing techniques including

\subsubsection{OpenCV image processing} We first utilize OpenCV library with traditional image processing techniques for head counting. To overcome the influence by moving objects such like a door, background subtraction method is fused with the color detection method in headcount detection. Unfortunately, we note that the accuracy of such method becomes unacceptable when the light inside the room is switched off. This motivates us to use deep learning techniques in the headcount camera. 

\subsubsection{Motivation for transfer learning}Recently, deep learning, especially, convolution neural network (CNN) has made great progress in  object recognition. However, building an object recognition model with CNN is tedious due to significant amount of data and resource requirement for training purpose. For instance, the ImageNet ILSVRC model was trained on $1.2$ million images over a period of $23$ weeks in multiple GPUs. As a consequence, it has become popular among researchers and practitioners to use  \emph{transfer learning} and fine-tuning, i.e., transferring the network weights trained on a rich dataset to the designated task, e.g., detecting people on images in this study. In particular, we use the pre-trained SSD Multibox model~\cite{Liu_ECCV_2016} as the network due to its higher accuracies, high frame rate, and suitability for embedded application. 

\subsubsection{SSD multibox}SSD matches objects with default boxes of multiple features maps with different aspects ratios. Each element of the feature map has either $4$ or $6$ number of default boxes associated with it. Any default box with a Jaccard overlap higher than a threshold of $0.5$ with a ground truth box is considered a match. SSD has six feature maps in total, each responsible for a different scale of objects, allowing it to identify objects across a large range of scales. SSD runs $3\times 3$ convolution filters on the feature map to classify and predict the offset to the default boxes. 

\subsubsection{Transfer learning and fine tuning}\label{sec:TLandFT} There are two main procedures that we adopt while using the pre-trained SSD Multibox model. First the \emph{transfer learning}. We use an SSD model that is pre-trained for COCO dataset \cite{Huang_CVPR_2017}. Then, we change the number of classes in box predicator according to our requirement.  Second, the fine-tuning. We keep the pre-trained weights of the feature extractor and use them as initial values for retraining. Note that since the feature extractor is trained for a large rich dataset over millions of iterations, the weights are stable and converged. As such, we only fine tune the feature extractor weights according to our dataset.\\

\begin{figure*}[t]
\centering {\includegraphics[scale=0.5]{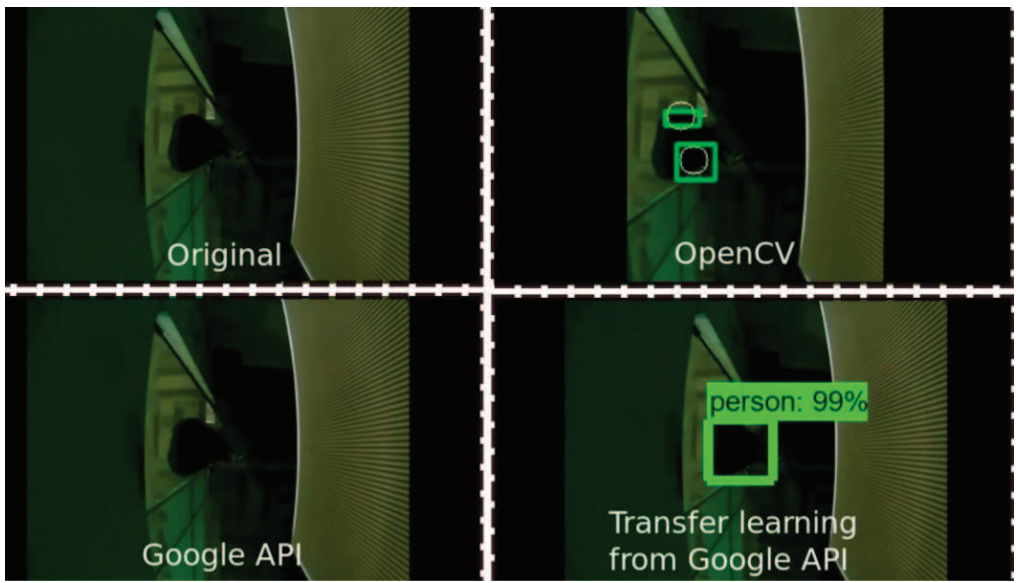}} 
\caption{Results of detecting using four different methods in low lighting conditions. Top Left: Original frame. Top Right: Using OpenCV libraries. Bottom Left: Using SSD model trained for COCO dataset. Bottom Right: Using Transfer Learning.}	 
\label{fig:fig4}
\end{figure*}

In Fig.~\ref{fig:fig4}, we depict the results by using OpenCV libraries using only the pre-trained SSD model and by using transfer learning. Fig.~\ref{fig:fig4} also demonstrates a frame that was captured in low lighting condition. It can be observed from the figure that OpenCV and pre-trained SSD model do not show good results under low lighting conditions. As shown in the top right corner of Fig.~\ref{fig:fig4}, the OpenCV method captures the same person as two instances and pre-trained SSD model does not capture anything at all (bottom left corner of Fig.~\ref{fig:fig4}). Nonetheless, the transfer learning method is well generalized for both good lighting conditions as well as low lighting conditions (bottom right corner of Fig.~\ref{fig:fig4}) and can be used for the people counting algorithm.

It is important to note that to reduce the processing complexity and power consumption for occupancy detection, MobileNet architecture is used as the feature extractor for the SSD model. The input image size was restricted to $250\times250$ pixels without loosing the accuracy, which greatly reduce the computational power. Further, while the original pre-trained SSD MobileNet can detect 99 object classes accurately, we reduce the number of classes to 2. This helps to reduce the power consumption of the module without limiting the performance of the detection (since our intention is only to detect person vs non-person case).

\subsection{Occupancy detection with sound sensor}Since a motion sensor has limited range, we also explore the suitability of using the sound sensor for occupancy detection. We are motivated to explore sound sensor due to the fact that vision approach with a camera can capture the identity of occupants and record the activities performed by the occupants within a selected space of the building. It not only violates the user's privacy but also need huge storage space and data rate for processing in real time. As such, existing studies on occupancy detection in building environment such as \cite{Jeon_BE_2018} and \cite{Candanedo_EB_2016} have used techniques that exploit environmental information including CO$_2$ level, noise level, humidity level, and particulate matter concentration to detect occupancy instead of using a camera. In this perspective, an acoustic approach using sound sensor is a potential alternative that is being explored in this work. The sound sensor is typically used for the detection of the loudness in ambient, where an input is taken from a microphone and amplified. We use a low-cost analog sound sensor, which has an SNR of $55$ dB at $1$ kHz  maximum input frequency level. Note that due to different room structure (room size, wall material, furniture), sensors are placed at different locations of different room, which results in a collection of different noise levels. The sensor data is sampled in every $100$ ms. Based on our empirical research and the previous study \cite{Billy_IoT_2017}, this sampling rate of $10$ Hz is enough to distinguish human activities around the environment. In addition, such low sampling would help reduce energy consumption and computation of sensors without losing necessary information. To accurately detect people occupancy in selected areas, we explore three different techniques including 1) threshold method, 2) unsupervised learning through clustering method, 3) supervised learning through deep learning method, and 4) a semi-supervised learning technique that uses the deep classifier to label the unsupervised results. 
\begin{figure*}[t]
\centering
\begin{minipage}{\linewidth}
\centering
\includegraphics[width=\textwidth]{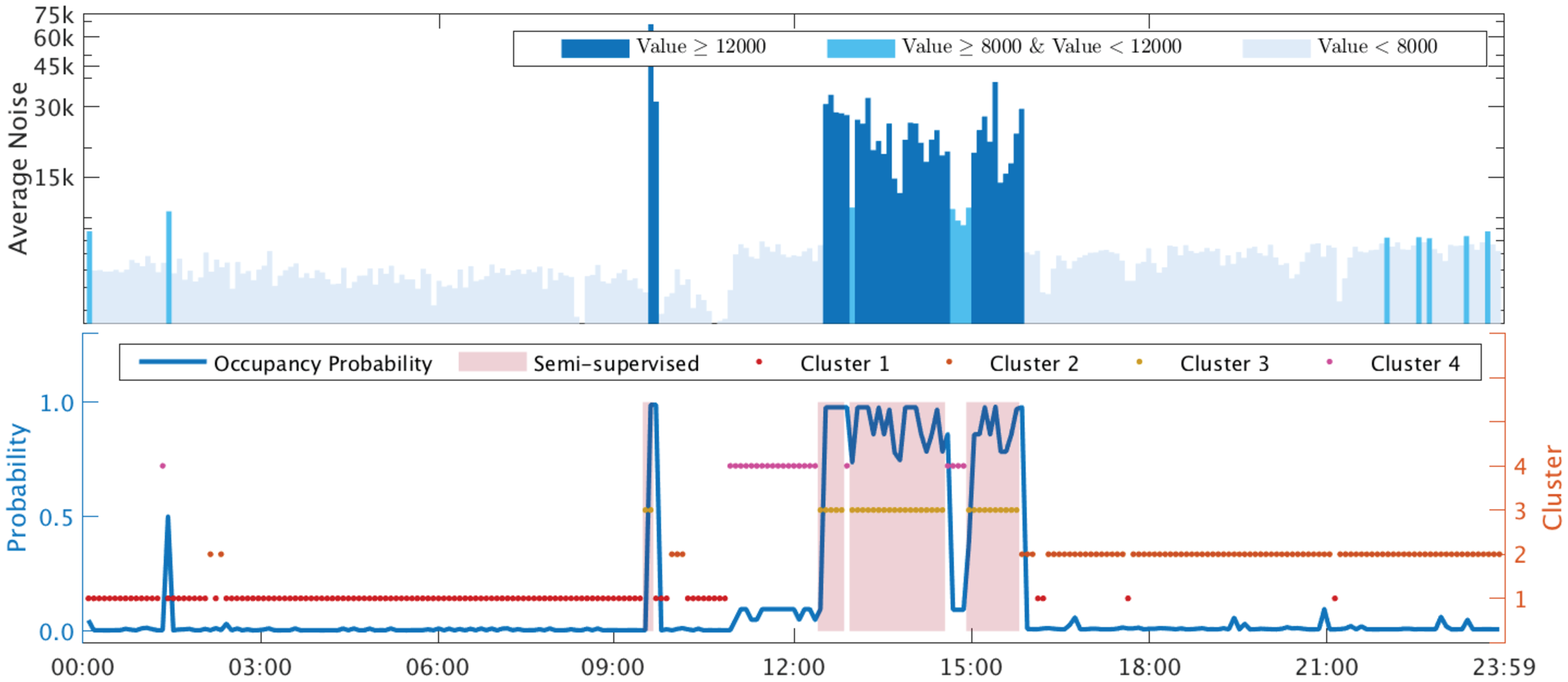}
\subcaption{\footnotesize{Meeting room P02.}}
\label{fig:fig2_a}
\end{minipage}
\\
\begin{minipage}{\linewidth}
\centering
\includegraphics[width=\textwidth]{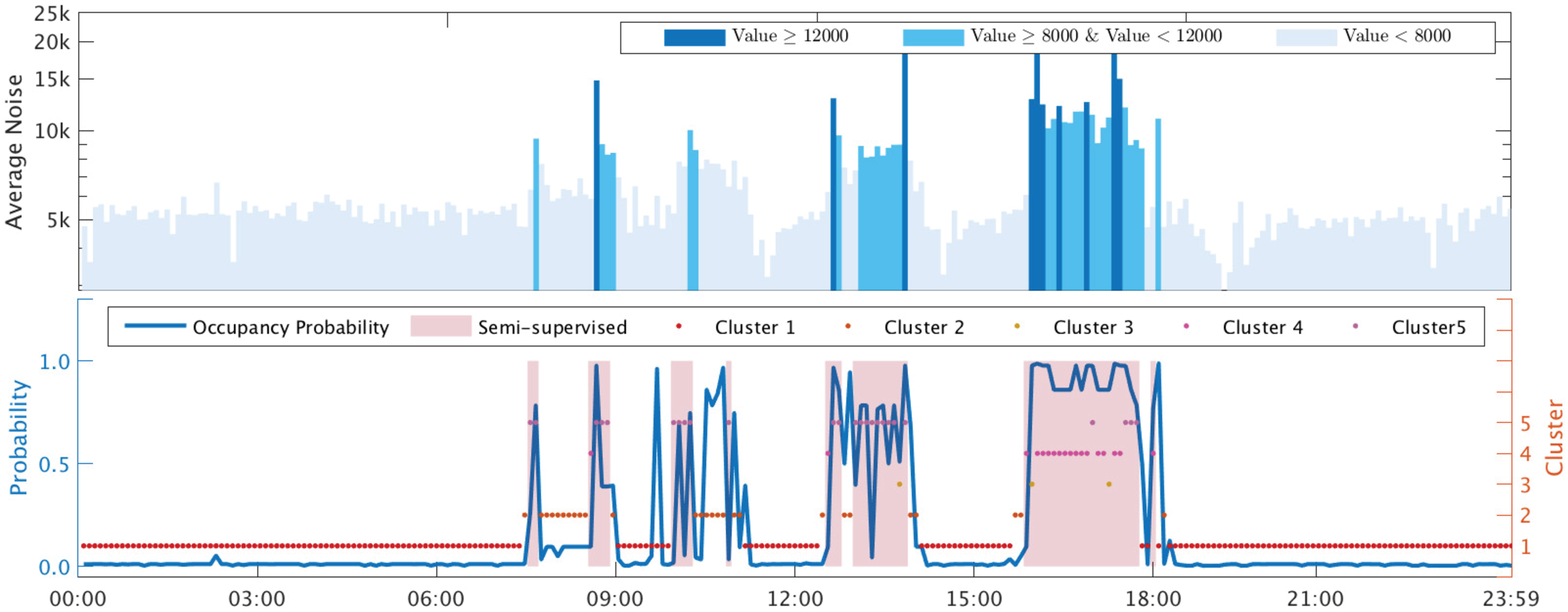}
\subcaption{\footnotesize{Meeting room P04.}}
\label{fig:fig2_b}
\end{minipage}
\caption{This figure demonstrates the results of occupancy detection in two different meeting rooms labeled as P02 and P04. We went to the actual site one day and recorded the ground truth for evaluation.}
\label{fig:ResultsOccDet}
\end{figure*}

\subsubsection{Thresholding method} We accumulate the sound sensor data in every $5$ minutes period and measure an empirical threshold value for each room in order to determine the occupancy. We observe that threshold values are different for different rooms and sometimes even for different days. As it can be seen Fig.~\ref{fig:ResultsOccDet} that the appropriate threshold for human activity detected for room P04 is $8000$, while the threshold becomes $12000$ for room P02. Therefore, just using the threshold method is not robust. Further, it is tedious to calibrate and finding the optimal threshold for each individual room.
\subsubsection{Unsupervised learning through clustering method}To have a robust occupancy detection through noise sensor without any calibration, we utilize unsupervised learning technique. Instead of just sending accumulated noise value, we send the noise histogram\footnote{The noise histogram refers to a set of bins (ranges), in which each bin represents a specific range of data collected by the sound sensors.} for each $5$ minutes interval according to the following arrangement: Bin 1 (sample values range from 0-6), Bin 2 (sample values range from 6-10), Bin 3 (sample values range from 10-15), Bin 4 (sample values range from 15-30), Bin 5 (sample values range from 30-50), Bin 6 (sample values range from 50-75), Bin 7 (sample values range from 75-100), and Bin 8 (sample values range from 100-above). While these $8$ levels of the histogram are used in our case, we believe $4$ or even fewer levels could also be sufficient. 

We adopt a similar unsupervised learning process in our previous work \cite{Billy_IoT_2017} in an outdoor smart city environment. In order to generate meaningful features of histogram data, we utilize the localized behavior of wavelet transformation, and Haar basis function\footnote{Haar basis function is a mathematical function that generates  the feature set for the histogram representation of data. In this work, Haar basis function is used as the mother wavelet due to its discontinuous nature.} is used as the mother wavelet due to its own discontinues nature. We represent each five-minute histograms in terms of Haar basis function. Further, principal component analysis (PCA) is used to remove the high correlation between histogram bins. Furthermore, we remove the redundant noise of data by only using the $n$ principal components that capture $95\%$ variation of data. Moreover, we use hierarchical clustering and calculate the optimal number of clusters using Calinski-Harabasz index. 
\subsubsection{Supervised learning through deep learning}A major issue of the unsupervised learning technique is that it is up to the human to interpret the outcome. For example, in our case, we may get $3$ to $5$ different clusters as the output of the unsupervised learning as shown in Fig~\ref{fig:ResultsOccDet} ($4$ clusters for room P02 and $5$ clusters for Room P04). In some cases, one cluster represents for the unoccupied duration, while in some other cases, two clusters may represent the unoccupied duration. Since the unsupervised learning method does not provide any meaningful understanding about the clusters, we use a deep neural network classifier, using the thresholding method with``reasonable threshold" as ``ground truth". The outputs of the classifier are the probability of occupancy. For instance, if the probability of the occupancy is greater than $0.5$, we determine the space as occupied.

To this end, first, we use a sparse auto-encoder to extract meaningful features for histograms. When training for feature extraction, we only use histograms that are related to one class of data, i.e. use only the histograms related to the occupied class, to train the auto-encoder rather than using the data of both classes, i.e., occupied and non-occupied. By doing so, we expect to construct more distinguishable features for histograms.  

When training the classifier, we fix the weights of the pre-trained auto-encoder and thus only optimize the weights of the deep neural network (DNN). To do so, we first use the sparse auto-encoder to construct sparse features for the histogram and then use these features as inputs to the deep classifier. In the training stage, we use a dataset of $5000$, half of which is used to train the auto-encoder. Training is done in $30000$ training epochs in both cases. As mentioned above, we only use a single class to train the auto-encoder. By doing so, we can improve the overall classification accuracy by around $3\%$ that compared to training using both the classes.
\subsubsection{Semi-supervised learning method}In order to improve the detection accuracy, we use the following semi-supervised learning technique. For a particular cluster from unsupervised learning, we look at their corresponding probability obtained by the deep classifier. If the majority say occupied, then we label that whole cluster as occupied or if the majority say unoccupied, then we label that whole cluster as unoccupied. Therefore, we obtain more consistent result compared using only the deep classifier.\\

Results of the aforementioned methods are depicted in Fig.~\ref{fig:ResultsOccDet} for two different meeting rooms, namely P02 and P04 respectively. It can be observed the optimal threshold values are different for different rooms (e.g., $12000$ for P02 and $800$ for P04), and those optimal threshold values are different for different rooms, and finding those values in large scale deployment could be tedious. Next, we demonstrate how the semi-supervised learning that combines both clustering and a deep classifier overcomes this problem. We train the deep classifier using data from another two meeting rooms (P01 and P06 with thresholds $11,000$ and $12,000$ respectively) and verify on P02 and P04, which is shown in  Fig.~\ref{fig:ResultsOccDet}. Finally, it can be seen that the semi-supervised learning method provides a robust and consistent occupancy detection on P02 and P04 even the training set is coming from P01 and P06.

% =======================================================
\subsection{Building efficiency via IoT based BMS}\label{sec:CaseStudy}
% =======================================================
\begin{figure}[t]
\centering
\includegraphics[width=0.8\columnwidth]{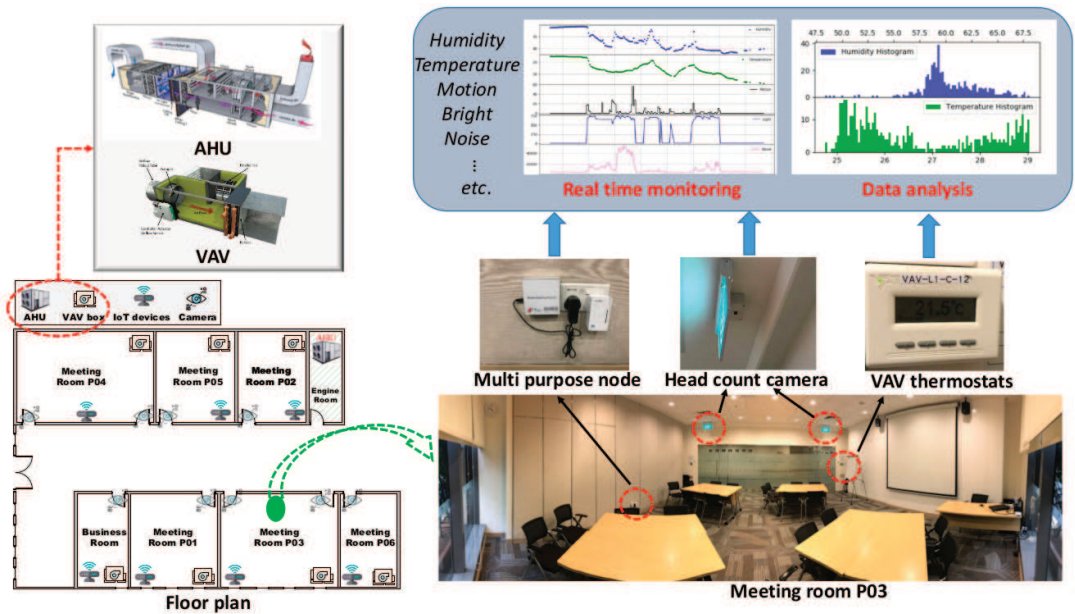}
\caption{Demonstration of the green building testbed.}
\label{fig:SelectedFloor}
\end{figure}

In this section, we demonstrate how our designed IoT based BMS can provide future insight into the building's efficiency of its HVAC system. The HVAC system is chosen particularly due to the fact that the HVAC system is responsible for more than $50\%$ energy consumption of commercial buildings.  A commercial facility in Singapore is considered as the green building testbed. In the selected building, the HVAC system is set up with modern chiller plant and central air handling unit (AHU) that can be managed by a typical BMS. The AHU contains an available speed supply fan, cooling coils, filters, mixing box, return air fan, dampers, and several variable air volume (VAV) terminal units that supply chilled air from the AHU to the terminal zones. We select a specific area of the commercial facility for experiment, which is meeting room P$03$ in Fig.~\ref{fig:SelectedFloor}. The schematic of the selected floor plan is shown in Fig.~\ref{fig:SelectedFloor}. In this floor, there are one AHU and several rooms with VAV units. Each room is equipped IoT sensors, and headcount cameras at the entrances. The multi-purpose node is responsible for collecting the surrounding environmental information. The environmental information that are collected for this experiment include temperature, humidity, light intensity, motion, and noise. On the other hand, headcount cameras are used to count people who enter and exit the rooms within the selected area. 

Note that the occupancy within a selected space has direct influence over the energy consumption of HVAC system, which in conjunction with information on the respective energy consumption pattern of the HVAC system can be exploited to regulate the energy consumption of HVAC. For example, in \cite{WenTai_TETC_2017}, the authors propose a data-driven approach for the energy consumption of HVAC system, in which the setpoint temperature of the HVAC is regulated according to the people activity and occupancy to control the energy consumption. Such control is shown to be very effective that can reduce an HVAC's energy consumption in a house by 33\%. An example of another study that utilizes people occupancy pattern for controlling the energy consumption of HVAC can be found in \cite{Aftab_EB_2017}.

We select one meeting room of the testbed to demonstrate the energy usage analysis. To do so, we extract historical data for the duration from November 01, 2017 to December 07, 2017. After filtering out the weekend days and days with missing data (due to maintenance and/or network fault), we have data of total $21$ days for analysis. The AHU operates from $7.40$ am to $7.40$ pm on each weekday, which is maintained by the building manager. However, we consider the office hours from $8.00$ am to $7.00$ pm for our study as the building is mostly occupied by people only during this period of time.

Based on the collected data on room temperature, chilled air temperature, and air flow, we compute the cooling energy for the energy usage analysis. In this context, the required cooling energy for a room is considered as $Q_\text{cooling} = \rho_\text{air}\mathcal{C}_\text{air}m_a(T_\text{room} - T_\text{supply})$~\cite{Balaji_2013}, where $\rho_\text{air}, \mathcal{C}_\text{air}, m_a, T_\text{room}$ and $T_\text{supply}$ represent  the air density at $20^\circ\mathrm{C}$, the specific heat of air, the volume of air supply, room temperature and chilled air temperature respectively. According to the $Q_\text{cooling}$ formula, the cooling energy consumption by a room relies on three variables; room temperature, chilled air temperature, and air flow, among them, chilled air temperature, and airflow is controlled by AHU and responsible for reducing room temperature to achieve the set point of room temperature, which is set to $20^\circ\mathrm{C}$ for the selected room. As such, the higher room temperature is, the more cooling energy is consumed.

\begin{figure}[t]
\centering
\includegraphics[width=0.6\columnwidth]{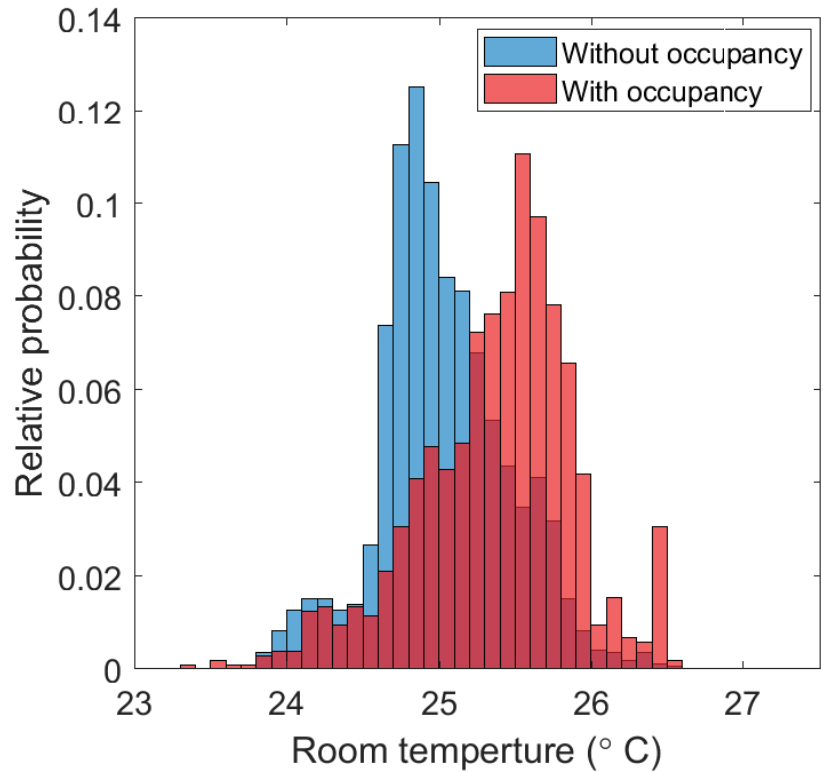}
\caption{Illustration of the distribution of room temperature with and without people occupancy.}
\label{fig:RoomTempDist}
\end{figure}

The air density and the specific heat coefficient are given as 1.204 kg/m$^3$ and 1.012 kJ$/^\circ\mathrm{C}/$kg respectively. Then, the data of room temperature is extracted for our IoT sensors. The supplied air volume and temperature are monitored by BMS, and hence the date of volume and temperature can be fetched from the database of BMS. We separate total time period of an experiment into multiple time slots where each time slot has 5 minutes time interval. Then, we classify the time period with occupancy and without occupancy, respectively, by utilizing the occupancy detection technique explained in Section~\ref{sec:MultiSensorFusion}. We obtain two sets of room temperature for time periods with and without people occupancy. Fig.~\ref{fig:RoomTempDist} shows the distribution of two sets of room temperature. Noticeably, the room temperature is higher when the room is occupied compared to the case without occupancy. Consequently, we infer that more cooling energy will be necessary to be consumed during  time periods with occupancy than that without any occupancy.
\begin{figure}[t]\centering
\includegraphics[width=0.6\columnwidth]{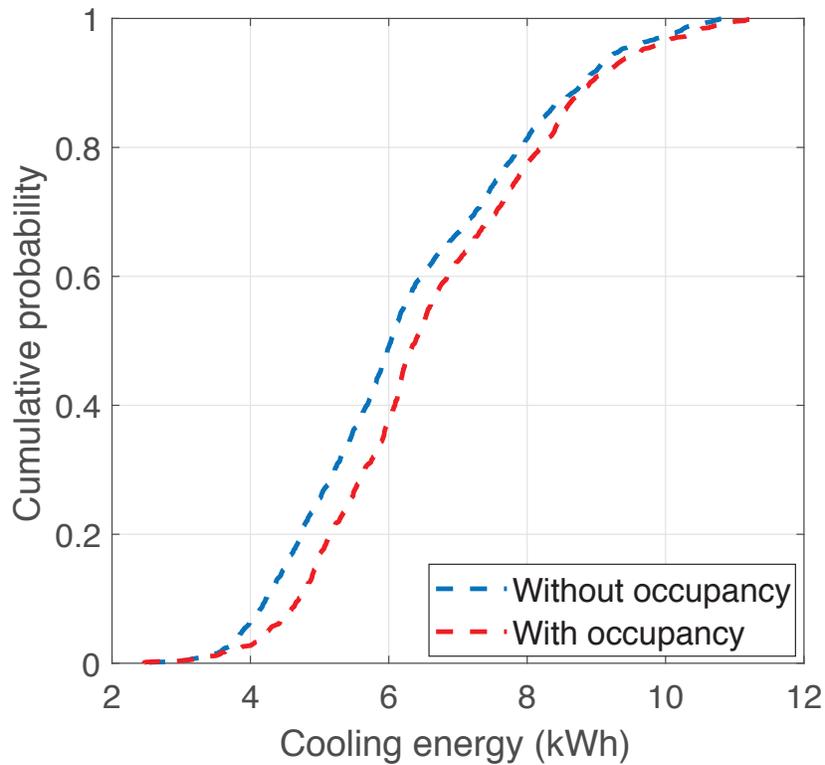}
\caption{Comparison of cooling energy usage with and without people occupancy.}
\label{figure:CoolingEnergyComparison}
\end{figure}

Further, we also classify the cooling energy usage based on the occupancy status. In Fig. \ref{figure:CoolingEnergyComparison}, we show the cumulative distributions of energy usages with and without people occupancy, and a gap between the cooling energy usages for the two considered occupancy status is clearly visible. Based on this figure, the average cooling energy consumption with occupancy and without occupancy in each time slot is $6.57$ kWh and $6.04$ kWh, respectively. Hence, the average cooling energy consumption with occupancy is $8.7\%$ higher than that without occupancy. Furthermore, the average room temperature with and without occupancy is $25.37^\circ\mathrm{C}$ and $25.02^\circ\mathrm{C}$, respectively, which indicates an increase of $0.35^\circ\mathrm{C}$ of average room temperature due to people occupancy.

Based on the above analysis, it is clear that energy consumption by the HVAC system of a building is highly dependent on its occupant status as it depends on other factors like building's thermal characteristics, usage of various appliances, the status of window blind and outdoor climate. Thus, this study provides useful insight on the energy efficiency of HVAC systems of a building by exploiting its people occupancy status, which can easily be achieved by simply deploying IoT devices in the building, as explained in this paper. For instance, the occupancy information can be used to develop suitable control strategies or optimization tools via signal processing techniques to opportunistically reduce the energy consumption by the HVAC and subsequently help the occupants to reduce their energy cost. For more information on how the occupancy information can be used to develop suitable control strategies or optimization tools please refer to \cite{Mirakhorli_EB_2016,Guo_LRT_2010,Yang_EB_2016}.

\section{Conclusion}\label{sec:Conclusion} In this paper, first, we have reviewed the application of IoT based signal processing techniques for managing various subsystems within a building. Then, we have provided an overview of how machine learning can be applied to the IoT device to detect people occupancy within a building that significantly contributes to the building's energy consumption. In particular, 1) a transfer learning based technique that counts people from images captured at the entrances of the selected areas of a building and 2) an unsupervised learning technique labeled by deep learning based occupancy detection that uses information obtained by sound sensors. Further, we have provided a short description of the testbed where these techniques are deployed for occupancy detection and shown how the information can help to gain insights on the use of HVAC system within the testbed. The correlation between occupancy and energy usage, as demonstrated in this study, has the potential to be used developing different energy management schemes that will help reducing energy consumption and electricity bills for the building manager.

There are many directions that the work reported in this paper can be extended. 
\subsubsection*{Widespread deployment}This proposed study conducted on a smaller scale of a real commercial building in Singapore. While using a real facility provides real user and system data for our study and analysis, it would be more interesting to see how the findings from the proposed study can be extended to a more larger scale, e.g., the entire building via widespread deployment of IoT devices. 
\subsubsection*{Detail modeling of energy usage}Another interesting extension of the proposed work would be to use machine learning techniques to perform a more detailed modeling of energy usage by various appliances of the building, and then design suitable techniques to optimize the usage for reducing electricity cost. 
\subsubsection*{Applying IoT beyond building energy} The IoT sensors also provide valuable data for predictive maintenance and anomaly detection as well. Hence it would be interesting to explore how  the collected data from the IoT sensors can be used helping a building to perform asset management, which may help a building to reduce its cost other than energy reduction. 
\subsubsection*{Exploration of quantitative performance} This paper presents qualitative results for the head counting and occupation detection study and their importance for the building management. However, there is a need to extend this work to present more quantitative analysis in terms of performance compared with existing studies in the literature and analysis of how the head counting impacts the energy consumption of the building.
\section*{Acknowledgement}This work was supported in part by the project NRF2015ENC-GBICRD001-028 funded by National Research Foundation (NRF) via the Green Buildings Innovation Cluster (GBIC), which is administered by Building and Construction Authority (BCA) - Green Building Innovation Cluster (GBIC) Program Office, in part by the SUTD-MIT International Design Centre (IDC; idc.sutd.edu.sg), in part by the grant NSFC 61750110529, and in part by the U.S. National Science Foundation under Grants CNS-1702808 and ECCS-1549881. Any findings, conclusions, or opinions expressed in this document are those of the authors and do not necessarily reflect the views of the sponsors.
%\def\baselinestretch{0.9}
%\bibliographystyle{IEEEtran}
%\bibliography{SPM_IoT}

\end{document}